\documentclass[twocolumn,showpacs,preprintnumbers,amsmath,amssymb]{revtex4}

\usepackage{epsfig,psfrag}

\newcommand{\T}{\mathcal{T}}

\begin{document}

\title{Hanbury-Brown-Twiss correlations and noise
in the\\ charge transfer statistics through a multiterminal Kondo
dot}
\author{T.~L.~Schmidt$^1$, A.~Komnik$^2$, and A.~O.~Gogolin,$^3$}
\affiliation{ ${}^1$ Physikalisches Institut,
Albert--Ludwigs--Universit\"at Freiburg, D--79104
Freiburg, Germany\\
${}^2$Service de Physique Th\'eorique, CEA Saclay, F–-91191
Gif-sur-Yvette, France \\
${}^3$Department of Mathematics, Imperial College London, 180
Queen's Gate, London SW7 2AZ, United Kingdom }
\date{\today}

\begin{abstract}
 We analyze the full counting statistics of charge transfer
 through a quantum dot in the Kondo regime, when coupled to
 an arbitrary number of terminals $N$.
 At the unitary Kondo fixed point and
 for $N>2$ we recover distinct anticorrelations of currents
 in concurring transport channels, which are related to the fermionic
 Hanbury Brown and Twiss (HBT) antibunching.
 This effect weakens as one moves away from the fixed point.
 Furthermore, we identify a special class of current correlations
 that are due entirely to the virtual polarization of the Kondo singlet.
 These can be used for extracting information on the parameters of
 the underlying Fermi-liquid model.
\end{abstract}

\pacs{72.10.Fk, 73.63.-b, 73.23.-b, 72.15.Qm}

\maketitle

Since its discovery the Kondo effect remains one of the most
active fields in condensed matter physics \cite{hewson}. The last
decade witnessed a very important achievement from the
experimental point of view: the creation of a fully adjustable
artificial Kondo impurity
\cite{goldhaber-gordon,kouwenhoven,weisvonklitzing}. This enables
one, in principle, to measure almost any transport property of the
system in a wide parameter range. One impressive example is the
Kondo resonance splitting, which occurs under nonequilibrium
conditions when the impurity is simultaneously coupled to two
terminals held at different chemical potentials
\cite{meirwingreenlee,koenigschoen,roschkroha}. It turned out that
the best way to explore this effect is to introduce a third
electrode and measure the tunnelling density of states
\cite{sunguo,lebanon}. This fact triggered a number of different
studies of the Kondo effect in multiterminal geometries
\cite{defranceschi,leturcq,sanchezlopez,roschshah}.

Apart from peak splitting measurements, artificial Kondo
impurities are also interesting for another reason. As was pointed
out in early works on transport in multiterminal systems, they
possess some unique transport properties due to a possibility of
having multiple concurrent transport channels. In particular, the
fermionic nature of charge carriers leads to antibunching of
individual transport events, also called fermionic Hanbury Brown
and Twiss (HBT) correlations \cite{buettiker,TMartin}. This
reveals itself as a negative sign of the current-current
correlation function between concurring transport channels. For
the simplest three-terminal geometry this effect has been
confirmed experimentally in \cite{heinzel,yamamoto}. Thus far
little has been done to investigate these effects in the Kondo
setup. The most advanced work is \cite{sanchezlopez}, where the
current-current correlations have been analysed for the
three-terminal geometry using the slave-boson approximation. To
the best of our knowledge, nothing is known about the correlations
of higher orders or close to the fixed point. With this Letter we
would like to close this gap by addressing the full counting
statistics (FCS) of such setups near the unitary limit.

We start with the following Hamiltonian,
\begin{eqnarray}                      \label{Ham0}
 H = \sum_{j=1}^N \, H_0[ \psi_{j \sigma}] + J \, {\bf S} \,
 \sum_{i,j=1}^N \sum_{\sigma \sigma'} \psi^\dag_{i \sigma}(0) \,
 \boldsymbol{\tau}_{\sigma \sigma'} \, \psi_{j \sigma'}(0)
 \nonumber \\
 - h \, S^z \, ,
\end{eqnarray}
where the first sum describes the dynamics of the electrons $H_0[
\psi_{j \sigma}] = \sum_{p \sigma j} (\epsilon_p - V_j)
\psi^\dag_{p \sigma j} \psi_{p \sigma j}$ in the noninteracting
leads (We set $e=\hbar=k_B=1$ throughout. The conductance quantum
then becomes $G_0 = 2e^2/h =1/\pi$). Here $\psi_{p\sigma j}$ is
the annihilation operator for an electron with momentum $p$, spin
$\sigma$ in the $j$th terminal held at chemical potential $V_j$.
The second part represents the coupling of the local spin degrees
of freedom in the leads to the impurity spin ${\bf S}$: $J$ is the
respective coupling constant, $\boldsymbol{\tau}$ denotes the set
of Pauli matrices, $h=\mu_B g B$ is the product of Bohr's magneton
$\mu_B$, of the gyromagnetic ratio $g$ and the magnetic field $B$
applied to the impurity. Via simple unitary transformation to a
set of new fermions $c_{j \sigma}$ this Hamiltonian can be mapped
onto the single channel Kondo model, where only one of the new
fields, $c_{1 \sigma}$, remains coupled to the impurity spin:
\begin{eqnarray}
 c_{1 \sigma} = \frac{1}{\sqrt{N}} \sum_{j=1}^N \psi_{j \sigma} \,.
\end{eqnarray}
In the low energy sector, well below the Kondo temperature $T_K$,
this fermion degree of freedom is hybridised with the impurity
spin into a singlet state. The true fixed point of the system is
just an $N$-leg resonant model. The effective fixed-point
Hamiltonian is simply $H_0$ for $c_{1\sigma}$, supplemented by
voltage terms. The best possible transparency is then realized by
the following scattering matrix $\psi_j(0^+) = s_{jk} \psi_k(0^-)$
($0^\pm$ denote in- and out-going plane waves), with \cite{nayak}
\begin{eqnarray}
 s_{jj} = (N-2)/N \, \, \, , \, \, \, s_{jk} = -2/N \, \, \,
 \mbox{for} \, \, \, j \neq k \, .
\end{eqnarray}

The central object in FCS calculations is the cumulant generating
function (CGF) $\chi(\boldsymbol{\lambda}) = \sum_{\boldsymbol{Q}}
e^{i {\bf Q} \boldsymbol{\lambda}} P({\bf Q})$ \cite{lll}, where
$\boldsymbol{\lambda}=(\lambda_1,...,\lambda_N)$ are the measuring
fields and $P({\bf Q)}$ is the probability that charges ${\bf
Q}=(Q_1,...,Q_N)$ are transferred across the respective channels
during the waiting time ${\cal T}$. The cumulants (irreducible
moments) are then found from the respective derivatives of
$\chi(\boldsymbol{\lambda})$. With the knowledge of the scattering
matrix (no interactions) the CGF is given by the multiterminal
version of the Levitov-Lesovik formula \cite{levitovlesovik}:
\begin{eqnarray}
 \ln \chi_0(\boldsymbol{\lambda}) = \frac{\cal T}{\pi} \int d
 \omega \, \ln \, \mbox{det} \left[ 1 +
 \hat{f}(\omega)(\hat{s}^\dag \, \widetilde{s} - 1 )\right] \, ,
\end{eqnarray}
where $\widetilde{s}_{mn} = e^{i (\lambda_n - \lambda_m)} \,
s_{mn}$ and the diagonal matrix $\hat{f}(\omega) = $diag$(n_1,
n_2,\dots,n_N)$ contains the Fermi distribution functions of
individual terminals $n_i$.
One then easily verifies that the fixed point CGF is given by:
\begin{eqnarray}
&& \ln \chi_0(\boldsymbol{\lambda}) = \frac{\cal T}{\pi}
\\  \nonumber  &\times&
  \int d \omega \, \ln \left\{
 1 + \left(\frac{2}{N}\right)^2 \sum_{i \neq j} \left[ e^{i
 (\lambda_j - \lambda_i)}-1\right] \, n_i \, (1 - n_j) \right\}
 \, .
\end{eqnarray}
In the particular case of zero temperature and three terminals such
that $V_1>V_2>V_3$ the above formula becomes
\begin{eqnarray}
 && \ln \chi_0 (\boldsymbol{\lambda})= \\ \nonumber
 && \frac{\cal T}{\pi} \left\{
 (V_1 - V_2) \ln \left[1+\frac{4}{9}\, \left( e^{i(\lambda_1 - \lambda_3)} + e^{i
 (\lambda_1 - \lambda_2)}- 2 \right) \right]
 \nonumber \right. \\ &+& \left. \nonumber
  (V_2 - V_3) \ln \left[ 1 + \frac{4}{9}\, \left( e^{i(\lambda_1 - \lambda_3)} + e^{i
 (\lambda_2 - \lambda_3)}- 2 \right) \right] \right\} \, .
\end{eqnarray}
Here one recognises that the current pairs $I_{1 \rightarrow 3}$
\& $I_{1 \rightarrow 2}$  and $I_{1 \rightarrow 3}$ \& $I_{2
\rightarrow 3}$ are mutually correlated, the corresponding
fluctuations of charge differences in the respective channels
being \cite{buettiker}
\begin{eqnarray}                 \label{Q2Q3}
 \langle \delta(Q_1- Q_2) \,  \delta(Q_1- Q_3) \rangle = -\frac{16}{81 \pi} \,
 {\cal T} \, (V_1 - V_2) \, , \nonumber \\
\langle \delta (Q_1 - Q_3) \, \delta (Q_2 - Q_3) \rangle =
-\frac{16}{81 \pi} \,
 {\cal T} \, (V_2 - V_3) \, .
\end{eqnarray}
The negative sign is the signature of fermion antibunching, or HBT
type correlations. On the other hand, the pair $I_{1 \rightarrow
2}$ \& $I_{2 \rightarrow 3}$ is not entangled in any way, as those
transport channels are not concurring. This is true only at $T=0$,
of course, its leading temperature dependence being
\begin{eqnarray}         \label{tempdep}
 \langle \delta (Q_1 - Q_2) \, \delta (Q_2- Q_3) \rangle = - \frac{16}{81 \pi} \,
 {\cal T} \, T \, .
\end{eqnarray}
for $T\ll |\mbox{min}(V_i-V_j)|$, $i\neq j$. In contrast to cross
correlations, the autocorrelation functions are positive, e.~g.
one finds for the current noise between the channels $2$ and $3$
\begin{eqnarray}               \label{auto}
 \langle  \delta^2 (Q_2 - Q_3) \rangle =
 \frac{20}{81} \, \frac{\cal T}{\pi} \, (V_2 - V_3) \,  .
\end{eqnarray}
For $V_2 = V_3$ the leading temperature dependence of this
correlation is again linear,
\begin{eqnarray}                      \label{auto1}
\langle  \delta^2 (Q_2 - Q_3) \rangle =
  \frac{8 }{9 \pi} \, {\cal T} \, T \,  .
\end{eqnarray}
One can show that these cumulants vanish whenever $V_2 = V_3$ and
$T=0$ even away from resonance and for a lead- asymmetric
junction.

The departure from the Kondo fixed point is described by the
Nozi\`{e}res Fermi-liquid theory \cite{nozieres}. The effective
Hamiltonian contains then in addition to the free part a
scattering term
\begin{eqnarray}
 H_{\mbox{sc}} = \frac{\alpha}{2 \pi \nu T_K} \sum_{p p' \sigma}
 (\epsilon_p + \epsilon_p') \, c^\dag_{1 p \sigma} c_{1 p' \sigma} \,
 ,
\end{eqnarray}
with a dimensionless amplitude $\alpha$ and local density of states
at the Fermi energy of the leads $\nu$ (assumed for simplicity to
be equal for all terminals), and an interaction:
\begin{eqnarray}
 H_{\mbox{int}} = \frac{\phi}{\pi \nu^2 T_K} \, ( c^\dag_{1 \uparrow} c_{1
 \uparrow} - n_0)( c^\dag_{1 \downarrow} c_{1
 \downarrow} - n_0) \, ,
\end{eqnarray}
with an amplitude $\phi$. Here $c_{1 \sigma} = \sum_p c_{1 p \sigma}$
and $n_0$ stands for the particle density in the zero field and
accounts for the background ion charge. The constant $\alpha$ is
responsible for the impurity specific heat whereas the sum $\alpha
+ \phi$ governs the magnetic susceptibility of the system. In the
actual Kondo model the Fermi-liquid relation $\alpha = \phi = 1$
holds \cite{nozieres}.

The standard way to calculate the FCS is to introduce
time-dependent counting fields $\boldsymbol{\lambda}$ (nonzero
only during the waiting time ${\cal T}$) coupled to charge
currents \cite{lll,nazarovlong}. In the alternative procedure such
terms can be gauged away by a canonical transformation
\cite{unitaryFCS}, which is easily generalized to the
multi-terminal geometry. This transformation is performed at the
expense of having $\lambda$ dependent exponentials embedded into
the interaction parts of the Hamiltonian via $H_{\mbox{sc}}[c_1] +
H_{\mbox{int}}[c_1] \rightarrow H_\lambda[c_1]$. The CGF is then
calculated according to the prescription \cite{reznikov},
\begin{eqnarray}                  \label{chi0}
 \chi(\boldsymbol{\lambda})&=& \chi_0(\boldsymbol{\lambda})
 \chi_1(\boldsymbol{\lambda})
 \nonumber \\
 &=& \chi_0(\boldsymbol{\lambda}) \,
 \left \langle T_C \exp \left[ - i \int_C d t \, H_\lambda(t) \right]
 \right\rangle  \, ,
\end{eqnarray}
where $C$ denotes the Keldysh contour and $T_C$ is the time
ordering operator on it \cite{AndersonFCS}. The evaluation of $
\chi_1(\boldsymbol{\lambda})$ can now be accomplished via the
standard linked cluster expansion in the couplings $\alpha,\phi$
using the Green's functions of the $\boldsymbol{\lambda}$-rotated
$c_{1 \sigma}$ operator:
\begin{eqnarray}\label{gfbare}
    \hat{g}_\lambda(\sigma, p, \omega)
=
    i \pi \, \frac{2}{N} \, \delta(\varepsilon_p-\omega) \sum_{j=1}^N
    \big\{ [n_{j \sigma}(\omega)-1/2] \, \boldsymbol{\tau}_0
\nonumber\\
    + \, e^{-i\lambda_j/2} \, n_{j \sigma}(\omega) \, \boldsymbol{\tau}_+
    -[1-n_{j \sigma} (\omega)] \, e^{i\lambda_j/2} \, \boldsymbol{\tau}_- \big\}\;,\nonumber
\end{eqnarray}
%\begin{eqnarray}\label{gfbare}
%    \hat{g}_\lambda(p,\omega)
%& = &
%    i \pi \, \frac{2}{N} \, \delta(\varepsilon_p-\omega)
%    \sum_{j=1}^N
%    \left(
%      \begin{array}{cc}
%        n_j-1/2 & e^{i\lambda_j/2} n_j \\
%        -e^{i\lambda_j/2} [1-n_j] & n_j-1/2 \\
%      \end{array}
%    \right)
%\end{eqnarray}
where $\boldsymbol{\tau}_\pm$ are the standard Pauli matrices,
$\boldsymbol{\tau}_0$ is the unity matrix and $n_{j \sigma}$ stand
for the Fermi distribution function in the respective terminal.

To the lowest (second) order there are three different
contributions: the ones proportional to $\alpha^2$, $\phi^2$, and
$\alpha \phi$. The latter cross term is nonzero only in finite
field. The $\alpha^2$ part emerges as
%\begin{widetext}
\begin{eqnarray}   \nonumber
 \delta_\alpha \ln \chi_1 = \left( \frac{2}{N}
 \right)^2 \frac{\alpha^2 {\cal T}}{2\pi T_K^2}
 \sum_{m \neq n}^N \sum_{\sigma}
 \left[ e^{-i
 (\lambda_m - \lambda_n)/2} - 1 \right] \,
 \\ \nonumber
 \times \, I_1(V_m + h \sigma,
 V_n + h \sigma)  \, ,
\end{eqnarray}
%\end{widetext}
where we define the function
\begin{eqnarray}
 I_1(V_1,V_2) = (V_1 - V_2) \frac{(\pi T)^2 + V_1^2 + V_2^2 + V_1
 V_2}{3 [ 1 - e^{-(V_1 - V_2)/T}]} \, .
\end{eqnarray}

Evaluation of the interaction $\phi^2$ correction results in two
different contributions. The first one is field independent and
for the case of $N=3$ terminals (a general result for arbitrary
$N$ exists but is quite lengthy)
%, for $N=2$ case see
%\cite{unitaryFCS})
is found to be
\begin{widetext}
\begin{eqnarray}\label{int1}
 & & \delta_\phi' \ln \chi_1 = \frac{1}{3^4}
  \frac{\phi^2 {\cal T}}{\pi T_K^2} \left\{ \sum_{m \neq n}^3
  \left[ \, 8 \, e^{-i (\lambda_m - \lambda_n)/2} \, I_2(V_m - V_n)
  + e^{-i (\lambda_m - \lambda_n)} \, I_2[2 (V_m - V_n)] \right]
 \right. \\
 & & \nonumber \left.
  +
  2 \sum_{m=1}^3 \left[\exp \left( - i \lambda_m + i/2 \sum_{n \neq m}^3
  \lambda_n \right) \, I_2 \left(2 V_m - \sum_{n \neq m}^3 V_n
  \right) + \exp \left(  i \lambda_m - i/2 \sum_{n \neq m}^3
  \lambda_n \right) \, I_2 \left(-2 V_m + \sum_{n \neq m}^3 V_n
  \right) \right]
  \right\} \, ,
\end{eqnarray}
\end{widetext}
where we have defined the function
\begin{eqnarray}
 I_2 (V) = \frac{V}{3} \frac{ 4(\pi T)^2 +
 V^2}{1-e^{-V/T}} \, .
  \nonumber
\end{eqnarray}
On the other hand, for the field-dependent correction we obtain
 \begin{eqnarray} \nonumber
 \delta''_{\phi} \ln \chi_1
= \left( \frac{2}{N} \right)^2 \frac{\phi^2 h^2 \T}{\pi T_K^2}
\sum_{m \neq n}^N [e^{-i(\lambda_m - \lambda_n)/2}-1] \nonumber \\
\times \frac{V_m-V_n}{1-e^{-(V_m-V_n)/T}}
 \, .
\end{eqnarray}
Finally, the $\alpha\phi$ cross term, which is also nonzero in a
finite field only, is given by
\begin{eqnarray}
 \delta_{\alpha \phi} \ln \chi_1
 &=& \left( \frac{2}{N} \right)^2 \frac{2\alpha \phi h^2 \T}{\pi
 T_K^2} \sum_{m \neq n}^N \left[ e^ {- i (\lambda_m - \lambda_n)/2}- 1
 \right]
 \nonumber \\
 &\times&
 \frac{V_m - V_n}{1-e^{-(V_m-V_n)/T}} \, .
\end{eqnarray}
%\begin{figure}
%\epsfig{file=OhrenDiagramm.eps,height=1.5cm}
%\caption[]{\label{diagram} Schematic representation of the
%interaction contribution in finite field.}
%\end{figure}

The full FCS contains all transport characteristics of the
system. Here we would like to concentrate on two rather
interesting issues. The first question we want to address is to
what extent are the HBT type correlations affected away from the
fixed point. In order to discuss this we again use the convention
$V_1> V_2 = V_3$. Then the transport channels $1 \rightarrow 2,
1\rightarrow 3$ become concurrent. The only term, still containing
both differences of counting fields $\lambda_1 - \lambda_{2,3}$,
can be found in Eq.~(\ref{int1}). The respective correction to
(\ref{Q2Q3}) at $T=0$ is given by
\begin{eqnarray}
 \langle \delta(Q_1- Q_2) \, \delta (Q_1- Q_3) \rangle = \frac{4}{3^5}
 \, \frac{\phi^2 \, {\cal T}}{\pi T_K^2} \, (V_1 - V_2)^3 \, .
\end{eqnarray}
We see that the residual Kondo interactions lead to a weakening of
the HBT signature. The remarkable fact is that this correction
only depends on $\phi$ as well as its cubic voltage behaviour (in
contrast to the linear one of the conventional HBT) offers an
opportunity to extract this effective parameter from the
experimental data. However, the original HBT background might
complicate the measurements.

Luckily, this difficulty can be avoided by measurement of
different correlation functions. As we have already mentioned
above, at the fixed point there is no entanglement between
nonconcurrent channels. This is different in the Kondo system. For
instance, under the above conditions there is a residual
correlation between the particle numbers in channels $1$ and $3$,
which is absent at the true unitary limit. Using the result
(\ref{int1}) we immediately verify that
\begin{eqnarray}
 \langle \delta ( Q_1 - Q_2) \, \delta (Q_2- Q_3) \rangle = - \frac{\phi^2 \,
 {\cal T}}{2 \cdot 3^5 \, \pi T_K^2} (V_1 - V_2)^3 \,  .
\end{eqnarray}
Comparing this equation with the temperature dependence of the
primary contribution ($\ref{tempdep}$) we see that the interplay
of nonequilibrium and polarization of the Kondo singlet amounts to
residual correlations  at $T=0$. Similar fate meets the
autocorrelation (noise) of the $I_{2 \rightarrow 3}$ current.
While it is zero in the unitary limit, Eq.~(\ref{auto}) for $V_2 =
V_3$, it acquires a finite value away from the fixed point,
\begin{eqnarray}\label{noise}
 \langle \delta^2 ( Q_2 - Q_3) \rangle = \frac{\phi^2 \,
 {\cal T}}{ 3^5 \, \pi T_K^2} (V_1 - V_2)^3 \,  .
\end{eqnarray}
Since in this limit there is no background from the
leading order, we expect that these types of
correlation would be a good means to access the parameter $\phi/T_K$
in future experiments. Interestingly, the finite magnetic field
does not affect these conclusions.

As the system under consideration is a Fermi liquid it is natural
to ask whether there is a possibility to use this fact in order to
derive a universal CGF valid also beyond the leading corrections
to the unitary limit. Indeed, in the case of two terminals Oguri
succeeded in deriving an expression for the current for a more
general Anderson impurity model for \emph{arbitrary} on-site
repulsion $U$ for small voltages \cite{ogurix}. In
\cite{AndersonFCS} a related formula for a CGF has been proposed,
valid around the unitary limit as well as for small $U$
\cite{unitaryFCS}. It turns out that a similar program can be
performed in the present multiterminal case. Indeed, the
cross-voltage derivatives, $\partial^2/\partial V_1 \partial V_2$,
of the bare Green's functions vanish. Therefore the method of
Ref.\cite{ogurix} applies directly. For the relative currents [the
actual currents in the respective channels being given by $I_i =
(1/N)\sum_j \, I_{i \rightarrow j}$] we find
\begin{eqnarray}\label{relcurmulti}
I_{ij}=4 \, \frac{V_i-V_j}{\pi
N}\left[1-\frac{2\chi_e^2+\chi_o^2}{6\Gamma_N^2}
(V_i^2+V_iV_j+V_j^2) \nonumber \right. \\ \left.
-\frac{3\chi_o^2}{2N\Gamma_N^2}\sum\limits_i V_i^2+... \right]\; ,
\end{eqnarray}
where $\Gamma_N = N \Gamma = N \, \pi \rho_0 \gamma^2$, $\gamma$
is the tunnelling amplitude coupling the Anderson impurity level
to the terminals with the local tunnelling density of states
$\rho_0$. The objects $\chi_{e,o}$ are the even/odd
susceptibilities which are correlations of impurity population
probabilities with the same/different spin orientations,
respectively. This result is valid for small voltages $V_i \ll
\Gamma_N$ and is exact in interaction as the $U$ dependence is now
contained in $\chi_{e,o}$. These are essentially equilibrium
quantities and can be found with other techniques, e.~g. by Bethe
ansatz. In the unitary limit the identification of coefficients is
very simple: $\chi_e = \Gamma_N \alpha /T_K$ and $\chi_o =
\Gamma_N \phi/T_K$. After that substitution one then recovers the
results found from the CGF (\ref{chi0}). We expect that also in
Eq.~(\ref{noise}) and similar formulas the parameter $\phi$ can be
substituted by the exact susceptibility $\chi_o$, even though we
do not have a formal proof of this statement.

To summarize, we report analytical results for the charge transfer
statistics for the Kondo impurity coupled to an arbitrary number
of terminals. In the Kondo limit, when bias voltage, magnetic
field and temperature are all smaller than the Kondo temperature,
we use a generalization of the Nozi\`eres Fermi-liquid theory to
derive a perturbative expansion for the FCS in leading irrelevant
operators. At the true fixed point we recover with our method the
Hanbury Brown and Twiss anticorrelations between concurrent
transport channels. While these turn out to weaken away from the
unitary limit,  new correlations emerge due to the virtual
polarization of the Kondo singlet. The most promising of these is
the low-$T$ noise in the $2-3$ channel of the $N=3$ junction at
$V_1>V_2=V_3$. Measurements of the latter would enable one to
directly access the effective parameters of the Fermi liquid model
in experiments similar to those of \cite{heinzel,yamamoto}.
%looks like agood place to stop
%We expect them to become feasible in the
%nearest future. Furthermore, we have expressed the relative
%currents through the exact susceptibilities for the case of more
%general multi-channel Anderson impurity model and made contact to
%the predictions of the CGF.

The authors participate in the European network DIENOW. TLS and AK
acknowledge the support by the DFG. AK is supported by the Feodor
Lynen program of the Alexander von Humboldt foundation.
\bibliography{HBTpaper}

\begin{thebibliography}{10}
\expandafter\ifx\csname bibnamefont\endcsname\relax
  \def\bibnamefont#1{#1}\fi
\expandafter\ifx\csname bibfnamefont\endcsname\relax
  \def\bibfnamefont#1{#1}\fi
\expandafter\ifx\csname url\endcsname\relax
  \def\url#1{\texttt{#1}}\fi
\expandafter\ifx\csname urlprefix\endcsname\relax\def\urlprefix{URL }\fi
\providecommand{\bibinfo}[2]{#2}
\providecommand{\eprint}[2][]{\url{#2}}

\bibitem{hewson}
\bibinfo{author}{\bibfnamefont{A.~C.} \bibnamefont{Hewson}},
  \emph{\bibinfo{title}{The Kondo Problem to Heavy Fermions}}
  (\bibinfo{publisher}{Cambridge University Press, Cambridge},
  \bibinfo{year}{1997}).

\bibitem{goldhaber-gordon}
\bibinfo{author}{\bibfnamefont{D.}~\bibnamefont{Goldhaber-Gordon}},
  \bibinfo{author}{\bibfnamefont{H.}~\bibnamefont{Shtrikman}},
  \bibinfo{author}{\bibfnamefont{D.}~\bibnamefont{Mahalu}},
  \bibinfo{author}{\bibfnamefont{D.}~\bibnamefont{Abusch-Magder}},
  \bibinfo{author}{\bibfnamefont{U.}~\bibnamefont{Meirav}}, \bibnamefont{and}
  \bibinfo{author}{\bibfnamefont{M.~A.} \bibnamefont{Kastner}},
  \bibinfo{journal}{Nature} \textbf{\bibinfo{volume}{391}},
  \bibinfo{pages}{156} (\bibinfo{year}{1998}).

\bibitem{kouwenhoven}
\bibinfo{author}{\bibfnamefont{S.~M.} \bibnamefont{Cronenwett}},
  \bibinfo{author}{\bibfnamefont{T.~H.} \bibnamefont{Oosterkamp}},
  \bibnamefont{and} \bibinfo{author}{\bibfnamefont{L.~P.}
  \bibnamefont{Kouwenhoven}}, \bibinfo{journal}{Science}
  \textbf{\bibinfo{volume}{281}}, \bibinfo{pages}{540} (\bibinfo{year}{1998}).

\bibitem{weisvonklitzing}
\bibinfo{author}{\bibfnamefont{J.}~\bibnamefont{Schmid}},
  \bibinfo{author}{\bibfnamefont{J.}~\bibnamefont{Weis}},
  \bibinfo{author}{\bibfnamefont{K.}~\bibnamefont{Eberl}}, \bibnamefont{and}
  \bibinfo{author}{\bibfnamefont{K.}~\bibnamefont{von Klitzing}},
  \bibinfo{journal}{Physica B} \textbf{\bibinfo{volume}{256-258}},
  \bibinfo{pages}{182} (\bibinfo{year}{1998}).

\bibitem{meirwingreenlee}
\bibinfo{author}{\bibfnamefont{Y.}~\bibnamefont{Meir}},
  \bibinfo{author}{\bibfnamefont{N.~S.} \bibnamefont{Wingreen}},
  \bibnamefont{and} \bibinfo{author}{\bibfnamefont{P.~A.} \bibnamefont{Lee}},
  \bibinfo{journal}{Phys.~Rev.~Lett.} \textbf{\bibinfo{volume}{70}},
  \bibinfo{pages}{2601} (\bibinfo{year}{1993}).

\bibitem{koenigschoen}
\bibinfo{author}{\bibfnamefont{J.}~\bibnamefont{K{\"o}nig}},
  \bibinfo{author}{\bibfnamefont{J.}~\bibnamefont{Schmid}},
  \bibinfo{author}{\bibfnamefont{H.}~\bibnamefont{Schoeller}},
  \bibnamefont{and}
  \bibinfo{author}{\bibfnamefont{G.}~\bibnamefont{Sch{\"o}n}},
  \bibinfo{journal}{Phys.~Rev.~B} \textbf{\bibinfo{volume}{54}},
  \bibinfo{pages}{16820} (\bibinfo{year}{1996}).

\bibitem{roschkroha}
\bibinfo{author}{\bibfnamefont{A.}~\bibnamefont{Rosch}},
  \bibinfo{author}{\bibfnamefont{J.}~\bibnamefont{Kroha}}, \bibnamefont{and}
  \bibinfo{author}{\bibfnamefont{P.}~\bibnamefont{W{\"o}lfle}},
  \bibinfo{journal}{Phys.~Rev.~Lett.} \textbf{\bibinfo{volume}{87}},
  \bibinfo{pages}{156802} (\bibinfo{year}{2001}).

\bibitem{sunguo}
\bibinfo{author}{\bibfnamefont{Q.-F.} \bibnamefont{Sun}} \bibnamefont{and}
  \bibinfo{author}{\bibfnamefont{H.}~\bibnamefont{Guo}},
  \bibinfo{journal}{Phys.~Rev.~B} \textbf{\bibinfo{volume}{64}},
  \bibinfo{pages}{153306} (\bibinfo{year}{2001}).

\bibitem{lebanon}
\bibinfo{author}{\bibfnamefont{E.}~\bibnamefont{Lebanon}} \bibnamefont{and}
  \bibinfo{author}{\bibfnamefont{A.}~\bibnamefont{Schiller}},
  \bibinfo{journal}{Phys.~Rev.~B} \textbf{\bibinfo{volume}{65}},
  \bibinfo{pages}{035308} (\bibinfo{year}{2001}).

\bibitem{defranceschi}
\bibinfo{author}{\bibfnamefont{S.}~\bibnamefont{{De Franceschi}}},
  \bibinfo{author}{\bibfnamefont{R.}~\bibnamefont{Hanson}},
  \bibinfo{author}{\bibfnamefont{W.~G.} \bibnamefont{{van der Wiel}}},
  \bibinfo{author}{\bibfnamefont{J.~M.} \bibnamefont{Elzerman}},
  \bibinfo{author}{\bibfnamefont{J.~J.} \bibnamefont{Wijpkema}},
  \bibinfo{author}{\bibfnamefont{T.}~\bibnamefont{Fujisawa}},
  \bibinfo{author}{\bibfnamefont{S.}~\bibnamefont{Tarucha}}, \bibnamefont{and}
  \bibinfo{author}{\bibfnamefont{L.~P.} \bibnamefont{Kouwenhoven}},
  \bibinfo{journal}{Phys.~Rev.~Lett.} \textbf{\bibinfo{volume}{89}},
  \bibinfo{pages}{156801} (\bibinfo{year}{2002}).

\bibitem{leturcq}
\bibinfo{author}{\bibfnamefont{R.}~\bibnamefont{Leturcq}},
  \bibinfo{author}{\bibfnamefont{L.}~\bibnamefont{Schmid}},
  \bibinfo{author}{\bibfnamefont{K.}~\bibnamefont{Ensslin}},
  \bibinfo{author}{\bibfnamefont{Y.}~\bibnamefont{Meir}},
  \bibinfo{author}{\bibfnamefont{D.~C.} \bibnamefont{Driscoll}},
  \bibnamefont{and} \bibinfo{author}{\bibfnamefont{A.~C.}
  \bibnamefont{Gossard}}, \bibinfo{journal}{Phys.~Rev.~Lett.}
  \textbf{\bibinfo{volume}{95}}, \bibinfo{pages}{126603}
  (\bibinfo{year}{2005}).

\bibitem{sanchezlopez}
\bibinfo{author}{\bibfnamefont{D.}~\bibnamefont{Sanchez}} \bibnamefont{and}
  \bibinfo{author}{\bibfnamefont{R.}~\bibnamefont{Lopez}},
  \bibinfo{journal}{Phys.~Rev.~B} \textbf{\bibinfo{volume}{71}},
  \bibinfo{pages}{035315} (\bibinfo{year}{2005}).

\bibitem{roschshah}
\bibinfo{author}{\bibfnamefont{N.}~\bibnamefont{Shah}} \bibnamefont{and}
  \bibinfo{author}{\bibfnamefont{A.}~\bibnamefont{Rosch}},
  \bibinfo{journal}{Phys.~Rev.~B} \textbf{\bibinfo{volume}{73}},
  \bibinfo{pages}{081309} (\bibinfo{year}{2006}).

\bibitem{buettiker}
\bibinfo{author}{\bibfnamefont{M.}~\bibnamefont{B{\"u}ttiker}},
  \bibinfo{journal}{Phys.~Rev.~B} \textbf{\bibinfo{volume}{46}},
  \bibinfo{pages}{12485} (\bibinfo{year}{1992}).

\bibitem{TMartin}
\bibinfo{author}{\bibfnamefont{T.}~\bibnamefont{Martin}}, in
  \emph{\bibinfo{booktitle}{Les Houches Session LXXXI}}, edited by
  \bibinfo{editor}{\bibnamefont{{H.~Bouchiat et al.}}}
  (\bibinfo{publisher}{Elsevier}, \bibinfo{year}{2005}).

\bibitem{yamamoto}
\bibinfo{author}{\bibfnamefont{W.~D.} \bibnamefont{Oliver}},
  \bibinfo{author}{\bibfnamefont{J.}~\bibnamefont{Kim}},
  \bibinfo{author}{\bibfnamefont{R.~C.} \bibnamefont{Liu}}, \bibnamefont{and}
  \bibinfo{author}{\bibfnamefont{Y.}~\bibnamefont{Yamamoto}},
  \bibinfo{journal}{Science} \textbf{\bibinfo{volume}{284}},
  \bibinfo{pages}{299} (\bibinfo{year}{1999}).

\bibitem{heinzel}
\bibinfo{author}{\bibfnamefont{M.}~\bibnamefont{Henny}},
  \bibinfo{author}{\bibfnamefont{S.}~\bibnamefont{Oberholzer}},
  \bibinfo{author}{\bibfnamefont{C.}~\bibnamefont{Strunk}},
  \bibinfo{author}{\bibfnamefont{T.}~\bibnamefont{Heinzel}},
  \bibinfo{author}{\bibfnamefont{K.}~\bibnamefont{Ensslin}},
  \bibinfo{author}{\bibfnamefont{M.}~\bibnamefont{Holland}}, \bibnamefont{and}
  \bibinfo{author}{\bibfnamefont{C.}~\bibnamefont{Sch{\"o}nenberger}},
  \bibinfo{journal}{Science} \textbf{\bibinfo{volume}{284}},
  \bibinfo{pages}{296} (\bibinfo{year}{1999}).

\bibitem{nayak}
\bibinfo{author}{\bibfnamefont{C.}~\bibnamefont{Nayak}},
  \bibinfo{author}{\bibfnamefont{M.~P.~A.} \bibnamefont{Fisher}},
  \bibinfo{author}{\bibfnamefont{A.~W.~W.} \bibnamefont{Ludwig}},
  \bibnamefont{and} \bibinfo{author}{\bibfnamefont{H.~H.} \bibnamefont{Lin}},
  \bibinfo{journal}{Phys.~Rev.~B} \textbf{\bibinfo{volume}{59}},
  \bibinfo{pages}{15694} (\bibinfo{year}{1999}).

\bibitem{lll}
\bibinfo{author}{\bibfnamefont{L.~S.} \bibnamefont{Levitov}},
  \bibinfo{author}{\bibfnamefont{W.~W.} \bibnamefont{Lee}}, \bibnamefont{and}
  \bibinfo{author}{\bibfnamefont{G.~B.} \bibnamefont{Lesovik}},
  \bibinfo{journal}{Journ. Math. Phys.} \textbf{\bibinfo{volume}{37}},
  \bibinfo{pages}{4845} (\bibinfo{year}{1996}).

\bibitem{levitovlesovik}
\bibinfo{author}{\bibfnamefont{L.~S.} \bibnamefont{Levitov}} \bibnamefont{and}
  \bibinfo{author}{\bibfnamefont{G.~B.} \bibnamefont{Lesovik}},
  \bibinfo{journal}{JETP Lett.} \textbf{\bibinfo{volume}{58}},
  \bibinfo{pages}{230} (\bibinfo{year}{1993}).

\bibitem{nozieres}
\bibinfo{author}{\bibfnamefont{P.}~\bibnamefont{Nozi{\`e}res}},
  \bibinfo{journal}{J.~Low~Temp.~Phys.} \textbf{\bibinfo{volume}{17}},
  \bibinfo{pages}{31} (\bibinfo{year}{1974}).

\bibitem{nazarovlong}
\bibinfo{author}{\bibfnamefont{Yu.~V.} \bibnamefont{Nazarov}},
  \bibinfo{journal}{Ann.~Phys. (Leipzig)} \textbf{\bibinfo{volume}{8
  (SI-193)}}, \bibinfo{pages}{507} (\bibinfo{year}{1999}).

\bibitem{unitaryFCS}
\bibinfo{author}{\bibfnamefont{A.~O.} \bibnamefont{Gogolin}} \bibnamefont{and}
  \bibinfo{author}{\bibfnamefont{A.}~\bibnamefont{Komnik}},
  \bibinfo{journal}{Phys.~Rev.~Lett.} \textbf{\bibinfo{volume}{97}},
  \bibinfo{pages}{016602} (\bibinfo{year}{2006}).

\bibitem{reznikov}
\bibinfo{author}{\bibfnamefont{L.~S.} \bibnamefont{Levitov}} \bibnamefont{and}
  \bibinfo{author}{\bibfnamefont{M.}~\bibnamefont{Reznikov}},
  \bibinfo{journal}{Phys. Rev. B} \textbf{\bibinfo{volume}{70}},
  \bibinfo{pages}{115305} (\bibinfo{year}{2004}).

\bibitem{AndersonFCS}
\bibinfo{author}{\bibfnamefont{A.~O.} \bibnamefont{Gogolin}} \bibnamefont{and}
  \bibinfo{author}{\bibfnamefont{A.}~\bibnamefont{Komnik}},
  \bibinfo{journal}{Phys.~Rev.~B} \textbf{\bibinfo{volume}{73}},
  \bibinfo{pages}{195301} (\bibinfo{year}{2006}).

\bibitem{ogurix}
\bibinfo{author}{\bibfnamefont{A.}~\bibnamefont{Oguri}},
  \bibinfo{journal}{Phys.~Rev.~B} \textbf{\bibinfo{volume}{64}},
  \bibinfo{pages}{153305} (\bibinfo{year}{2001}).

\end{thebibliography}

\end{document}